\documentstyle[epsfig]{mn}
\begin{document}


\title
[The temperatures of dust-enshrouded AGNs]
{The temperatures of dust-enshrouded active galactic nuclei}
\author[Neil Trentham \& A.\,W. Blain] 
{
Neil Trentham and A.\,W. Blain\\
Institute of Astronomy, Madingley Road, Cambridge, CB3 0HA.\\
}
\maketitle 

\begin{abstract} 
A high density of massive dark objects (MDOs),
probably supermassive black holes, in the centres of nearby galaxies has been 
inferred from recent observations.
There are various indications that much of the accretion responsible
for producing these objects took place in dust-enshrouded 
active galactic nuclei (AGNs). 
If so, then measurements of the
intensity of background radiation and the source counts in the far-infrared 
and submillimetre wavebands constrain the 
temperature of dust in these AGNs.  An additional 
constraint comes from the hard X-ray background, if this is produced by
accretion.
One possibility is that the dust shrouds surrounding the accreting AGNs
are cold, about $30\,{\rm K}$. 
In this event, the dusty AGNs could be some subset of
the population of luminous distant
sources discovered at $850\,\mu$m using the SCUBA array on the JCMT,
as proposed by Almaini, Lawrence and Boyle (1999). 
An alternative is that the dust shrouds surrounding the accreting AGNs are
much hotter ($> 60\,{\rm K}$).  
These values are closer to the dust temperatures
of a number of well-studied low-redshift ultraluminous galaxies that 
are thought to derive
their power from accretion.  If the local MDO density
is close to the maximum permitted,
then cold sources
cannot produce this density without the submillimetre
background being overproduced if they accrete at high radiative
efficiency, and thus 
a hot population is required.
If the
dust-enshrouded accretion occured
at similar redshifts to that taking place in unobscured optical quasars, 
then a significant fraction
of the far-infrared background radiation measured by 
$COBE$ at $140\,\mu$m, but very little of the submilllimetre background 
at $850\,\mu$m, may have been produced by 
hot dust-enshrouded AGNs which 
may have already been seen in 
recent {\it Chandra}
X-ray surveys.  

\end{abstract} 


\section{Introduction} 

There is increasing evidence 
(Magorrian et al.\ 1998; van der Marel 1999, Ferrarese \& Merritt 2000,
Gebhardt et al.~2000)
for massive dark objects (MDOs), probably 
supermassive black holes
(Kormendy \& Bender 1999), in the centers of most large nearby galaxies. 
The comoving mass density 
of these MDOs
is higher 
by at least a factor of two (Salucci et al.~1999), and possibly more
(Phinney 1997; Haehnelt, Natarajan \& Rees 1998), 
than can be explained by accretion onto 
optically-luminous AGNs.  

Haehnelt et al.\ (1998) suggested one possible solution to this problem: 
that much of the accretion 
is obscured by dust, and that this population of 
dust-obscured AGNs is missing from optically selected samples (see also
Fabian 1999 for a specific model).
Dust-enshrouded AGNs are certainly known at high redshift: see Table 1
for some examples, and McMahon et al.\ (1999) and Benford
et al.\ (1999).
However, note that the fraction of their bolometric luminosity that
originates from an AGN as opposed to a starburst has not been 
well established in most cases. 
In an important recent development, analysis of 
the hard X-ray background 
(Fabian \& Iwasawa 1999; Salucci et al.\ 1999) has revealed that if the
sources responsible for producing the X-ray background are obscured 
(type-2) quasars, then
they may also be the class of sources that are responsible
for producing the high local MDO density, as dust and the associated
density of molecular and neutral gas is transparent to hard 
X-rays.

\begin{table*}
\caption{Examples of dust-enshrouded high-redshift AGNs.} 
{\vskip 0.75mm}
{$$\vbox{
\halign {\hfil #\hfil && \quad \hfil #\hfil \cr
\noalign{\hrule \medskip} 
Object  & Redshift  & Reference  &\cr
\noalign{\smallskip \hrule \smallskip} 
\cr 
IRAS F10214+4724 & 2.3 & Lacy et al.\ (1998) and references therein &\cr 
QSO H1413+117       & 2.6 & Kneib, Alloin \& Pello (1998) and references 
         therein  &\cr 
SMM J02399$-$0136 & 2.8 & Ivison et al.\ (1998), Frayer et al.\ (1998) &\cr 
APM 08279+5255 & 3.9 & Irwin et al.\ (1998), Lewis et al.\ (1998) &\cr
BR 1202$-$0725 & 4.7 & Isaak et al.\ (1998), Kawabe et al.\ (1999) &\cr
SMM J02399$-$0134 & 1.1 & Soucail et al.~(1999), Bautz et al.~(2000) &\cr 
\noalign{\smallskip \hrule} 
\noalign{\smallskip}\cr}}$$}
\end{table*} 

Fabian \& Iwasawa (1999) suggest that the bulk of the 
energy absorbed at optical, ultraviolet and
soft X-ray
wavelengths is reradiated at
far-infrared and submillimetre wavelengths by obscured AGNs.  We 
investigate this 
scenario using constraints from far-infrared and submillimetre
source counts and backgrounds. 
If we assume that interstellar dust 
is responsible for the absorption and reradiation, then   
the question becomes: what is
the dust temperature of these sources?

Almaini et al.\ (1999) and Gunn \& Shanks (2000) recently proposed that
the obscured quasars contributing to the X-ray background have cold
($\leq 40$\,K) dust shrouds, and that they account for about
ten to twenty per cent of the mJy sources detected at
850\,$\mu$m using the SCUBA camera at the JCMT 
(Blain et al.\ 2000 and references therein).  
Their motivation for suggesting this temperature comes
from observations of 
high-redshift 
radio-quiet AGNs 
(Benford et al.\ 1999). 
Interestingly this is the most favoured temperature for the 
SCUBA sources derived from 
current measurements of the far-infrared and submillimetre 
background, source counts 
and redshift distribution (Blain et al.\ 1999a,b; Trentham, 
Blain \& Goldader 1999).  

This idea is very attractive. The dust in type-2 AGNs, which are required to 
produce the hard X-ray background, is observed directly by SCUBA
reradiating away the absorbed ultraviolet and 
soft X-ray energy in the submillimetre waveband.
Two of these SCUBA sources, in the field of the lensing cluster Abell 370
(Smail, Ivison \& Blain 1997), which are known from optical spectroscopy
to contain AGNs, have in fact been seen at X-ray
wavelengths with {\it Chandra} by Bautz et al.~(2000). 
On the other hand, much of the accretion may take place surrounded by
hotter ($ \geq 60$\,K) dust shrouds, as would be required by radiative
transfer models of the spectral energy distributions 
(SEDs) of luminous infrared
galaxies (Rowan-Robinson 2000 and references therein).  
These objects would then on the
whole have SEDs skewed to wavelengths
too short to be seen easily 
with SCUBA, yet too long to have been seen in ground-based
near-infrared surveys.  They may be present in substantial numbers
but until recently have been very difficult to detect at all.  
The first such objects to be seen in a systematic survey were recently
described by Wilman, Fabian \& Gandhi, who found $ISO$ 6.7-$\mu${\rm m}
and 15-$\mu${\rm m} counterparts of two hard X-ray sources detected
by {\it Chandra}. 
Many local
infrared-luminous AGNs, for example 
the `Warm ULIGs' of Sanders \& Mirabel (1996), 
have temperatures close to 60\,K. Higher temperatures 
in excess of 100\,K are observed for a 
few, rare high-redshift
ULIGs, such as P09104+4109 (Kleinmann et al.\ 1988) and APM 08279+5255
(Lewis et al.\ 1998), with temperatures in excess of 100\,K. 
The fortunate circumstances under which APM 08279+5255 (Irwin et al.~1998)
was discovered
(the infrared source happens to lie close to the caustic of a
lensing galaxy so is highly magnified) are perhaps suggestive that many
such objects (without such a large magnification) exist
but are as yet unidentified.
Even higher dust 
temperatures (150--200\,K) are implied by the observed bumps in the SEDs
of the most luminous Seyfert galaxies at about 25\,$\mu$m (Sanders \&
Mirabel 1996),
although the bolometric luminosities of these galaxies
are not dominated by these hot components.

Therefore it is 
worth investigating the constraints on the temperatures of 
dust-enshrouded AGNs, the observations of which are described in
Section\,2.
In Section\,3 we demonstrate the effect 
of the dust temperature on both the observable source counts and 
the efficiency of accretion. We conduct more detailed 
calculations in Section\,4. In Section\,5 we discuss our results 
and indicate the most promising lines of enquiry for the future.  
Throughout this paper we assume that 
$H_0 = 50$\,km\,s$^{-1}$\,Mpc$^{-1}$ and
$\Omega_0 = 1$.

\section{Dust-enshrouded AGNs}

It is plausible that dust-enshrouded AGNs are responsible for the production 
of the hard X-ray background radiation and constitute a subset of the SCUBA 
galaxies. There is an additional argument to support the presence of obscured 
AGNs. If the SCUBA sources have similar gas density structures 
as local ultraluminous infrared galaxies (ULIGs), specifically that the
bolometric luminosity originates from a very dense compact central
region (Downes \& Solomon 1998; Sakamoto et al.\ 1999), 
as suggested by their similar SEDs (Iviso et al.\
1998, 2000), then a very efficient power source is required. 
The only $z=0$ stellar systems that exist at 
the same densities, 
100\,M$_{\odot}$\,pc$^{-3}$ or greater, are the cores of elliptical galaxies
(Kormendy \& Sanders 1992); 
however, because only about one per cent of the 
cosmological density in stars (Fukugita, Hogan \& Peebles 1998)
is in these systems, 
it is unlikely that their formation alone 
could have generated the large fraction of the bolometric energy density 
of the Universe ($> 10$ per cent) that is accounted for by the SCUBA 
sources (Trentham 2000). 
High-mass stars would be another alternative, but this would then require
the MDOs to be clusters of stellar remnants, not black holes, 
an idea which seems
to be ruled out given the high angular resolution of
recent stellar-kinematical measurements of the centers of
nearby galaxies (e.g.~Kormendy
\& Bender 1999) unless the clusters of stellar remnants coalesce to form
supermassive black holes.

An alternative efficient power source would be accreting AGNs. 
However, it is 
unlikely that sources with bolometric luminosities that originate from these 
compact central regions, with cold rotating nuclear gas disks on 
scales $<300$\,pc,
can be powered by AGNs alone, if their only source of
fuel is these gas disks. This is because it is 
difficult to remove sufficient angular momentum from the gas to 
allow it to 
reach the event horizon of a black hole without converting the 
gas into stars (Begelman 1994). This is a long-standing astrophysical 
problem, and 
it is now confirmed by observations 
on scales where the 
central black hole contributes siginificantly to the local 
gravitational potential (e.g.~Sakamoto et al.\ 1999). 

\section{The observational effects of different dust temperatures 
in accreting AGNs}

A luminosity function can be adopted to describe the population of 
submillimetre-luminous galaxies, for example the Gaussian function chosen 
by Trentham et al.\ (1999).  This
can be integrated over redshift and luminosity 
to make predictions of the source counts and background radiation 
intensities in the submillimetre and far-infrared wavebands, if an 
(SED) is assumed, 
(which is normalised to the 
spectral luminosity at an arbitrary wavelength of 60\,$\mu$m, $L_{60}$, 
chosen for convenient comparison with {\it IRAS} studies of the local 
Universe.  This luminosity function
is consistent with the observed source counts of galaxies at 
850 and 175\,$\mu$m and the intensity of the far-infrared
background radiation 
(Puget et al.~1996,
Hauser et al.\ 1998; Schlegel et al.\ 1998, Finkbeiner et al.~2000) 
if the dust 
temperature $T \simeq 40$\,K, assuming that the dusty 
soures all radiate as modified blackbodies with a single temperature, 
and that the dust emissivity varies with frequency 
as a power law $\nu^{p}$ where $p=1.5$.

In Fig.\,1, we show the effects of changing the dust temperature, and thus 
modifying the SED, on the counts of galaxies at 850 and 175\,$\mu$m. The 
results are presented in the form of ratios 
as compared with the values at $T = 40$\,K, and so 
details of the normalization of the counts and backgrounds need not be 
considered here. The Gaussian luminosity function used has a 
characteristic 60-$\mu$m luminosity $L^{*}_{60} = 3 \times 10^{13}$ L$_{\odot}$ 
and a width $\sigma=0.5$. No luminosity evolution is assumed and the 
comoving density of sources is assumed to be constant in the redshift range 
$1 < z < 3$ and zero elsewhere (Trentham et al.\ 1999).

At 850\,$\mu$m the counts are less at greater dust temperatures, 
because the hotter sources are being probed further down the 
Rayleigh--Jeans tail of their SEDs, and so 
only sources with greater bolometric luminosity can exceed the flux 
density limit required to appear in the count. The temperature 
dependence of the 175-$\mu$m source counts, and of the background 
radiation intensity at 140 and 240\,$\mu$m, is much weaker as these 
quantities are derived from the properties of the SED much closer to its peak. 

If we assume that the bolometric far-infrared luminosity of the AGN, 
$L_{\rm bol}$, is powered entirely by accretion onto a black hole at a fraction
$f_{\rm edd}$ of the Eddington luminosity and that the accretion efficiency 
is independent of luminosity, then the mass of the central black 
hole (the MDO) is 
\begin{equation}
{ M \over {{\rm M}_{\odot}}} = 3.5 \times 10^{-5} {f_{\rm edd}}^{-1}
\left({ L_{\rm bol} \over {{\rm L}_{\odot}}}\right). 
\end{equation}
In this accretion model, a greater bolometric luminosity must derive 
from a greater accretion rate, and so correspond to a greater black hole 
mass. As shown in Fig.\,1, the comoving MDO density generated in order
to power the 850- and 175-$\mu$m sources depends on the dust
temperature, in the sense that very cold and very hot sources have greater
bolometric luminosities per unit spectral luminosity at the wavelength
of 60\,$\mu$m, at which the luminosity function is defined, or indeed per
unit bolometric luminosity.

\begin{figure}
\begin{center} 
\vskip-2mm
\epsfig{file=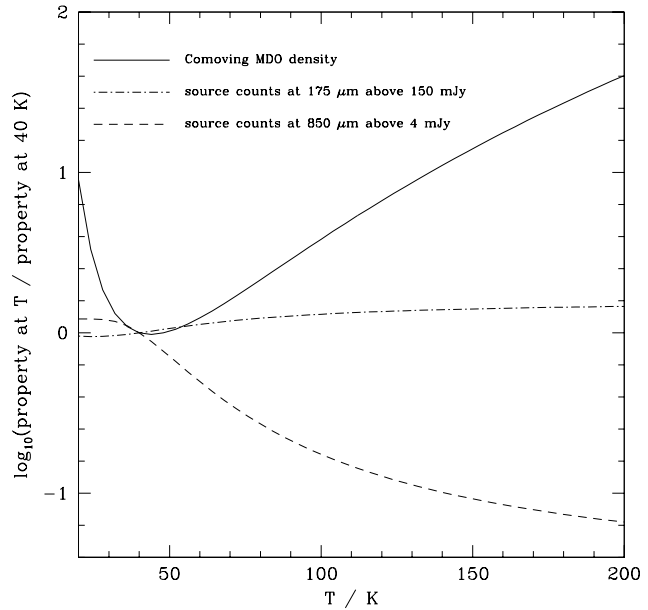, width=8.65cm}
\end{center}
\vskip-4mm
\caption{
The ratio of the comoving density of MDOs onto which material is 
accreting in dust-enshrouded AGNs, the
175-$\mu$m source counts brighter than 150 mJy and 
the 850-$\mu$m source counts brighter than 4 mJy, as a function of the 
dust temperature $T$ of the sources. The galaxies are all assumed to have 
a single value of $T$.  
The results are presented as ratios, and so the 
general form of the curves is insensitive to the details of the luminosity 
function parameters. 
}
\end{figure}

It is clear from Fig.\,1 that a greater comoving MDO density is produced 
by hotter sources per unit contribution to the 850-$\mu$m source counts.
If we assume that the bolometric far-infrared luminosities 
of the sources are proportional to their absorbed 
soft X-ray flux densities, that is 
the fraction of the energy absorbed by dust and gas
at wavelengths outside the 
soft X-ray 
window is either negligible or independent of luminosity, then the absolute 
value of the 
hard X-ray background intensity is almost proportional to the 
comoving MDO density. The difficulties in normalizing this relation
are discussed in Section 4. The dust 
temperature could thus
have a significant effect on the submillimetre/far-infrared
observability of the 
sources responsible for the generation of the hard X-ray background. 
Because the numbers plotted in Fig.\,1 are ratios, when
the models are matched to observations in detail, the absolute values 
of these (and other) quantities must also be considered as a function
of temperature.

\section{Specific normalized models of accretion} 

We now investigate the normalization of specific models of 
dust-enshrouded accretion onto black
holes at high redshift.  As discussed above, we can integrate over 
redshift and luminosity to derive the comoving mass density of the 
black holes that are accreting at any redshift $z$ using equation (1). 
By including the density of black holes that have accreted in 
the past, but are quiescent at $z$, we can compute the total mass 
density of MDOs at $z$.

We can constrain the redshift-dependant luminosity function of
$\phi (L_{60},z)$ of accreting sources, their temperatures $T$, 
and $f_{\rm edd}$  
in combination if we know the mass density of MDOs at $z=0$. Further 
constraints are imposed by measurements of the counts and backgrounds
if we assume an SED for the sources, and by the X-ray background 
if we include some infrared-to-X-ray conversion factor. 
A number of assumptions must first be made: 
\vskip 1pt \noindent
1) All sources are assumed to have the same temperature. 
In reality sources probably have a range of temperatures; consequently $T$ 
should be regarded as a luminosity-weighted average temperature.
\vskip 1pt \noindent
2) A specific form of $\phi(L_{60},z)$ must be 
assumed. The dependence on $L_{60}$ is assumed to be a Gaussian 
function with width $\sigma = 0.5$ and characteristic luminosity 
$L_{60}^{*}$. No luminosity evolution is assumed, that is $L_{60}^{*}$ 
does not vary with redshift, and the evolution of the comoving space 
density of sources is chosen to ensure that the integral 
$\int_{0}^{\infty} L_{60} \phi(L_{60},z) {\rm d}L$ varies with redshift in the 
same way as the ultraviolet AGN luminosity density (Boyle \& Terlevich 
1998).
\vskip 1pt \noindent 
3) Any mode of accretion that is either not dust-enshrouded or not 
generating powerful emission, for example advection-dominated
accretion, 
is assumed to be negligible.  
\vskip 1pt \noindent
4) When constraining $\phi (L,z)$, $T$, and $f_{\rm edd}$ using 
measurements of the backgrounds and source counts at 
various wavelengths, we require that the backgrounds or 
source counts are never overproduced. It is not a concern if the model 
underpredicts the observed quantities, as there is certainly a
significant contribution from infrared-luminous galaxies that are powered 
by star-formation activity as opposed to AGN accretion. 
\vskip 1pt \noindent
The models are constrained by the upper limit to the 2.8-mm source counts
(Wilner \& Wright 1997), the 850-$\mu$m and 450-$\mu$m source 
counts (Blain et al.\ 2000), the 175-$\mu$m source counts 
(Kawara et al.\ 1998; Puget et al.\ 1999), the submillimetre and far-infrared 
background radiation intensity (Puget et al.~1996,
Fixsen et al.\ 1998; Hauser et al.~1998;
Schlegel et al.\ 1998, Finkbeiner et al.~2000) 
and the MDO density at $z=0$, $\rho_{\rm MDO}(0)$. 
The methodology used to make these comparisons is described in detail by 
Trentham et al.\ (1999). 

\section{Results and discussion}

In all the models, very low dust 
temperatures are ruled out to avoid breaching both 
the limit to the observed 2.8-mm counts (Wilner \& Wright 1997) 
and the submillimetre background intensity (Fixsen et al.\ 1998). 
For most values of $L_{60}^{*}$, 
the most important constraint becomes the requirement that 
the 175-$\mu$m source counts are not overproduced, a constraint which
also depends on $T$. 
The 850- and 450-$\mu$m counts are exceeded for some 
values of $L^*$ and $T$, but these values are always also 
ruled out by the observed intensity of either the 850- or 240-$\mu$m 
backgrounds.  As the value of $L^{*}$ increases, the count constraints 
become steadily more important relative to the background constraints, 
but never become more significant for the values of 
$L_{60}^{*} < 10^{14}$\,L$_{\odot}$. 

The results can be 
summarized by a simple equation: 
\begin{equation}
{{t} \over {\rm K} }  \geq \left[
{{\rho_{\rm MDO}(0)}\over{{\rm M}_{\odot} {\rm Mpc}^{-3}}} \,  f_{\rm edd}
\right]^{0.36}.  
\end{equation}
This equation is an equality if the
submillimetre and far-infrared backgrounds are produced entirely 
by dust-enshrouded AGNs; it is an inequality if a significant fraction of 
the background comes from another population of objects. 

From equation (2), $T \geq 70$\,K is required for $f_{\rm edd} > 0.1$, 
if we assume the higher value of the local MDO density implied by the 
MDO mass--galaxy luminosity correlation of Magorrian et al.~(1998)
i.e.~$\rho_{\rm MDO}(0) = 1.5 \times 10^6 {\rm M}_{\odot} {\rm Mpc}^{-3}$. 
Such objects are probably not the SCUBA-selected galaxies, which 
probably have temperatures of about 40\,K (Ivison et al.\ 1998, 2000; 
Blain et al.\ 1999b; Trentham et al.\ 1999). An additional warm population 
of dusty galaxies must be hypothesized to explain the X-ray background 
and the local MDO density in this case. 

This scenario is interesting, because the observed 140- and 240-$\mu$m 
background radiation intensity can be explained 
entirely by this population, as shown in Fig.\,2. The cooler SCUBA sources 
in the models of Blain et al.\ (1999a,b) and Trentham et al.\ (1999) tend 
to underpredict the 140-$\mu$m background
and the 100-$\mu$m and 60-$\mu$m backgrounds
of Finkbeiner et al.~(2000).  Hence, a hotter additional 
population could be argued for {\it a priori}, and it is attractive that such a 
population can simultaneously provide the accretion necessary to 
generate the MDO density at $z=0$.

\begin{figure}
\begin{center}
\epsfig{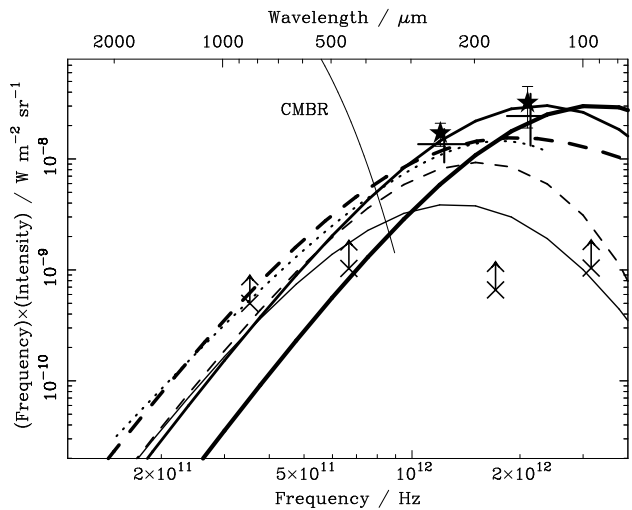}
\end{center}
\caption{The intensity of background radiation in the millimetre, submillimetre
and far-infrared wavebands, as deduced by: 
Fixsen et al.\ (1998) --
thin dotted line that ends
within the frame; Schlegel et al.\ (1998) -- stars; Hauser et al.\ (1998) 
-- thick solid crosses. The diagonal crosses
represent lower limits  
to the background intensity inferred from source counts. 
From left to right:
the 850- and 450-$\mu$m count of Blain et al.\ (2000), and the 175- and 
95-$\mu$m counts of Kawara et al.\ (1998) and Puget et al.\ (1999).
The three solid lines represent models for dust-enshrouded AGNs 
presented here: $T=30$\,K, $\rho_{\rm MDO} = 2 \times 10^5 
{\rm M}_{\odot}\,{\rm Mpc}^{-3}$, $f_{\rm edd} = 0.1$ -- (thin);
$T=70$\,K, $\rho_{\rm MDO} = 1.5 \times 10^6 {\rm M}_{\odot}\,{\rm Mpc}^{-3}$,
$f_{\rm edd} = 0.1$ -- (medium);
$T=100$\,K, $\rho_{\rm MDO} = 1.5 \times 10^6 {\rm M}_{\odot}\,{\rm Mpc}^{-3}$,
$f_{\rm edd} = 0.1$ -- (thick).
The thick solid line also fits the recently-determined
100-$\mu$m and 60-$\mu$m backgrounds
(Finkbeiner et al.~2000).
For comparison, the background spectrum in the Modified Gaussian
model of the SCUBA sources and backgrounds (Barger et al.\ 1999; 
Blain et al.\ 1999a) is represented by the thick dashed line, and 
that in model G of Trentham et al.\ (1999) is shown by the thin dashed line. 
There is a population of low-redshift luminous objects in the Modified 
Gaussian model, but not in model~G.
}
\end{figure}

If, on the other hand, we assume the lower value of the local MDO 
density implied by van der Marel (1999)
i.e.~$\rho_{\rm MDO}(0) = 2 \times 10^5 {\rm M}_{\odot} {\rm Mpc}^{-3}$, 
and that MDOs formed in 
only the dust-enshrouded high-redshift AGNs that contribute to 
the hard X-ray background, then sources with temperatures
as low as 30\,K are consistent with the observations. Such objects 
could easily make up a subset of the SCUBA sources. This is essentially
the scenario proposed by Almaini et al.~(1999) and studied 
by Gunn \& Shanks (2000).  

\subsection{Constraints from X-ray observations} 

It is difficult to use measurements of the hard X-ray 
background as a strong constraint on the history of accretion, 
because we do not know the SED of a typical obscured AGN over 
the entire spectrum from submillimetre to hard X-ray wavelengths. 
For example, the dust-enshrouded AGNs listed in Table 1 
have very different SEDs. 

Fabian \& Iwasawa (1999) estimated that the total X-ray flux density that 
is absorbed and reradiated by the dust-enshrouded AGN population is
about
3\,nW\,m$^{-2}$\,sr$^{-1}$. This is the integrated value over all 
wavelengths and is thus independent of the dust temperature. 
This is 58, 7, and 7 per cent of 
the background radiation integrated over all wavelengths for the three 
solid lines on Fig.\,2 in order of increasing thickness, and so in these 
three models, 42, 93, and 93 per cent repsectively
of the energy absorbed by the 
dust shrouds of the AGNs must come 
from either a population of Compton-thick 
sources that are not present in X-ray surveys, or from outside the 
X-ray window defined by Fabian \& Iwasawa (1999). 

One possibility is that very luminous heavily obscured AGNs, perhaps like
Markarian 273 (that looks like a starburst at near-infrared and optical
wavelengths but like a dusty AGN at mid-infrared wavelengths; 
Lutz et al.~1999), or maybe even more obscured,
are common at high redshift.  These could
well be Compton-thick.  
This possibility is suggested by the non-detection of six SCUBA sources 
with the $Chandra$ X-ray satellite (Fabian 
et al.~2000; 
see Table 2
and the accompanying list of energy budgets).
Such a scenario also would be suggested by the
spectroscopy of the SCUBA sources by Barger et al.~(1999), who 
find no Seyfert 1's like Markarian 231, so that if their sources
are AGN-powered they {\it must} be heavily obscured. 

Table 2 should be considered in conjunction with the following energy
contributions to the extragalactic background light. 
$\epsilon_{\rm cc}+\epsilon_{\rm hc} = 2 \,{\rm nW}\,{\rm m}^{-2}\
{\rm sr}^{-1}$
from
Fabian \& Iwasawa (1999), after removing their correction for Compton-thick
sources.
$\epsilon_{\rm cc}+\epsilon_{\rm cC} \leq 7 \, {\rm nW}\,{\rm m}^{-2}\,
{\rm sr}^{-1}$
from Trentham et al.~(1999), summing all the energy in the SCUBA sources, 
the temperatures 
of which are constrained to not overproduce the far-infrared
background at the high-$T$ end, and to produce about one-third of the
450-$\mu$m background at the low-$T$ end (Blain et al.~1999a).  
This is a lower limit due to the possibility of a contribution from sources
powered by star formation.
$\epsilon_{\rm cc}+\epsilon_{\rm hc} + \epsilon_{\rm cC}+\epsilon_{\rm hC}
= 5 \, {\rm nW}\,{\rm m}^{-2} \,
{\rm sr}^{-1}$ for the thin solid line
in Fig.~2.

\begin{table*}
\caption{Relative numbers of dust-enshrouded AGNs.  The terms ``hot" and
``cold" refer to objects with dust temperatures above and below 50 K
respectively.  The numbers 
listed in the fourth,
fifth, sixth, and seventh
columns are from the survey of Fabian et al.~(2000; F00),
Barger et al.~(2000;
B00), 
Hornschemeier et al.~(2000; H00), 
and Bautz et al.~(2000; B00a) respectively.
The detectability of sources stated in the second and third columns
adopts thresholds appropriate to these surveys. 
A direct comparison of the numbers within the
fourth through seventh columns is difficult since 
they depend on the sensitivities of {\it Chandra} and SCUBA in 
each survey.  The large
numbers in the first and fourth lines relative to the numbers in the
second line might be due to selection effects although the
numbers certainly suggest that the {\it Chandra} and SCUBA galaxies  
are quite distinct at the high-flux end.  
In addition Wilman et al.~(2000) reported two {\it Chandra} objects 
with $ISO$ 
detections at 6.7 $\mu$m and 15 $\mu$m that were not detected with SCUBA.  
The values for cold Compton-thick sources are upper limits due to possible
contributions from sources that are powered by star formation.  
Blank entries represent regimes not probed by
the particular surveys. 
}
{\vskip 0.75mm}
{$$\vbox{
\halign {\hfil #\hfil && \quad \hfil #\hfil \cr
\noalign{\hrule \medskip}
Type & SCUBA?  & $Chandra$?  &N$_{\rm F00}$
& N$_{\rm B00}$ & N$_{\rm H00}$ & N$_{\rm B00a}$ & Energy generated &\cr
\noalign{\smallskip \hrule \smallskip}
\cr
Cold Compton-thick & yes & no & $\leq 6$ 
& $-^{}$  & $\leq 10$ & $\leq 1$ & $\epsilon_{\rm cC}$ &\cr 
Cold Compton-thin & yes & yes & 1  & 
 1 & 0 & 2 & $\epsilon_{\rm cc}$ &\cr
Hot Compton-thick & no & no & $-^{}$ & $-^{}$ & $-^{}$ 
        & $-^{}$ &$\epsilon_{\rm hC}$ &\cr
Hot Compton-thin & no & yes & 3 & 
19 & 6 & $-^{}$ & $\epsilon_{\rm hc}$ &\cr
\noalign{\smallskip \hrule}
\noalign{\smallskip}\cr}}$$}
\end{table*}

The possibility of there existing many hot Compton-thick sources 
is also realistic as both more cold sources appear to be
Compton-thick than Compton-thin (see the first two lines of Table 2), 
and more Compton-thin sources appear to
be hot than cold (see the second and fourth lines of Table 2). 
Note that caution is required when directly interpreting
the numbers in the fourth, fifth and sixth columns of Table 2 since
the sensitivities of {\it Chandra} and SCUBA to the same
template object in each survey are different).  
Such sources may appear at hard X-ray wavelengths, like those probed 
more sensitively by the {\it XMM-Newton} satellite.   

If the entire observed intensity of the 140-$\mu$m background radiation, 
about 30\,nW\,m$^{-2}$\,sr$^{-1}$, 
ten times greater than
Fabian \& Iwasawa (1999) estimated, is to be explained by dust-enshrouded 
AGNs, then higher-temperature sources radiating at a high efficiency 
are required. 
However, other mechanisms, for 
example warm dust reradiating absorbed starlight from normal 
star-forming galaxies at $z \leq 2$ (see Appendix A), or an isotropic warm 
Galactic dust component (Lagache et al.\ 1999), could be responsible 
for a significant fraction of the 140-$\mu$m background, thus 
reducing the background radiation intensity that is required from this 
additional AGN population. 

In summary, we cannot distinguish between these scenarios at present. 
The properties of high-redshift dust-enshrouded AGNs remain only 
poorly constrained.  Nevertheless using the results presented here we can 
suggest productive lines of enquiry for the future. 

\subsection{Making observational 
progress in understanding the growth of MDOs and dust-enshrouded AGNs} 

More information about both the properties of distant dusty galaxies 
and the density of MDOs at low redshifts is required to better 
understand the dominant mode of AGN accretion at high redshifts. 
Based on the models preesented here, we now list the most 
promising directions for future progress.
\vskip 1pt
\noindent (1) Multiwaveband follow-up observations of 
high-redshift far-infrared and 
submillimetre selected galaxies will be required to identify candidate 
dusty AGNs that are undergoing significant amounts of 
accretion.  Optical and near-infrared
spectroscopy is currently available for some 
SCUBA sources (Ivison et al.~1998, 2000; Barger et al.~1999). 
This can be used to provide some 
evidence for the presence or absence of an AGN, but not  
to unambiguously determine the nature of the 
power source. The 
likely extreme optical 
faintness of some SCUBA galaxies (Barger, Cowie
\& Richards 2000; Smail et al.\ 2000) will make
spectroscopic observations of complete samples very difficult. 
Mid-infrared measurements of PAH features (for example Genzel 
et al.\ 1998) and the shapes and ratios of near-infrared emission lines 
(for example Veilleux et al.\ 1999) may allow the unambiguous diagnosis of 
AGNs using future instruments like $SIRTF$, $FIRST$ and $SPECS$.
\vskip 1pt
\noindent (2) 
The best wavelength at which to make an efficient search for more 
examples of dust-obscured AGNs depends on both the dust temperatures of 
the objects, and the sensitivity of detectors. The SED of an object with 
a dust temperature $T$ at redshift $z$ peaks at a wavelength of about 
$65 (1+z) (T/{\rm 40 \,K})^{-1}$\,$\mu$m in the observers frame. Hence,
cooler objects are probably best searched for at submillimetre wavelengths,
while warmer objects will probably be detected more efficiently in far-infrared 
surveys.  The properties of galaxies 
detected using SCUBA at 850-$\mu$m, {\it ISO} at 175-$\mu$m, and
IRAM at 1.25 mm (e.g.~Bertoldi et al.~2000) are 
already revealing the best strategies for future surveys (Blain 1999).
\vskip 1pt
\noindent (3) Improved modeling of the velocity dispersion anisotropy in
observations of the stellar kinematics in nearby galaxies (see Section 4.6 of 
van der Marel 1999) will hopefully allow a more precise determination of 
MDO masses, and thus of the local black hole density $\rho_{\rm MDO}(0)$. 
The improved form of the relation between galaxy luminosity and 
MDO mass that will result may also provide another constraint on the 
accretion history, since the local early-type galaxy luminosity function is
well known (e.g.~Binggeli, Sandage \& Tammann 1988). 
Correlating MDO masses with other properties, like velocity dispersion
(Ferrarese \& Merritt 2000; Gebhardt et al.~2000) will help too, particularly
if the correlation is tight.
\vskip 1pt
\noindent (4) More sensitive measurements of the hard ($\geq$ 10 keV) X-ray 
source counts will soon be available from 
{\it XMM-Newton}. 
By cross correlating 
sensitive hard-X-ray maps with deep submillimetre and 
far-infrared observations,
the nature of dust-enshrouded AGNs can probably 
be pieced together slowly on an object by object basis. 

Ultimately, these pieces of information will allow a 
more rigorous calculation 
similar to that described in Section 4 to carried out. This will allow the 
nature of the dust-enshrouded AGN population to be clarified, providing 
a strong constraint on $f_{\rm edd}$ and thus telling us about the 
fundamental physics of accretion.
 
\section{Conclusions}  
 
It is not yet possible to determine a unique description of the  
SEDs of the  
dust-enshrouded AGNs that are hypothesized to be responsible for both  
the hard X-ray background radiation intensity and the density of  
massive dark objects (MDOs) in the cores of low-redshift galaxies. 
If the density of MDOs is at the bottom of the current observational  
range, then the  
low-temperature dust-enshrouded AGNs described by Almaini et al.\  
(1999), a subset of the high-redshift SCUBA galaxies, can  
account for all the accretion. However, a hotter population of  
dust-enshrouded AGNs is required if the MDO density is towards the top  
of the range permitted by current observations. This second scenario  
has the advantage of accounting for  
most of the far-infrared background radiation.  
New clues have been provided by the lack of
strong cross-correlation between 
SCUBA and $Chandra$ sources, implying
that most cold sources, if they are AGNs, are
Compton-thick.  There are more hot than cold
Compton-thin sources, and so 
one might expect there to be significant numbers of
hot Compton-thick sources seen by neither $Chandra$ nor SCUBA.
A hotter population of dust-enshrouded AGNs does indeed seem to be a
realistic proposition.  
Deeper observations at higher X-ray energies with 
{\it XMM - Newton} will help to uncover such a population, if it
exists.

\begin{figure*}
\begin{minipage}{170mm}
{\vskip-3.5cm}
\begin{center}
\epsfig{file=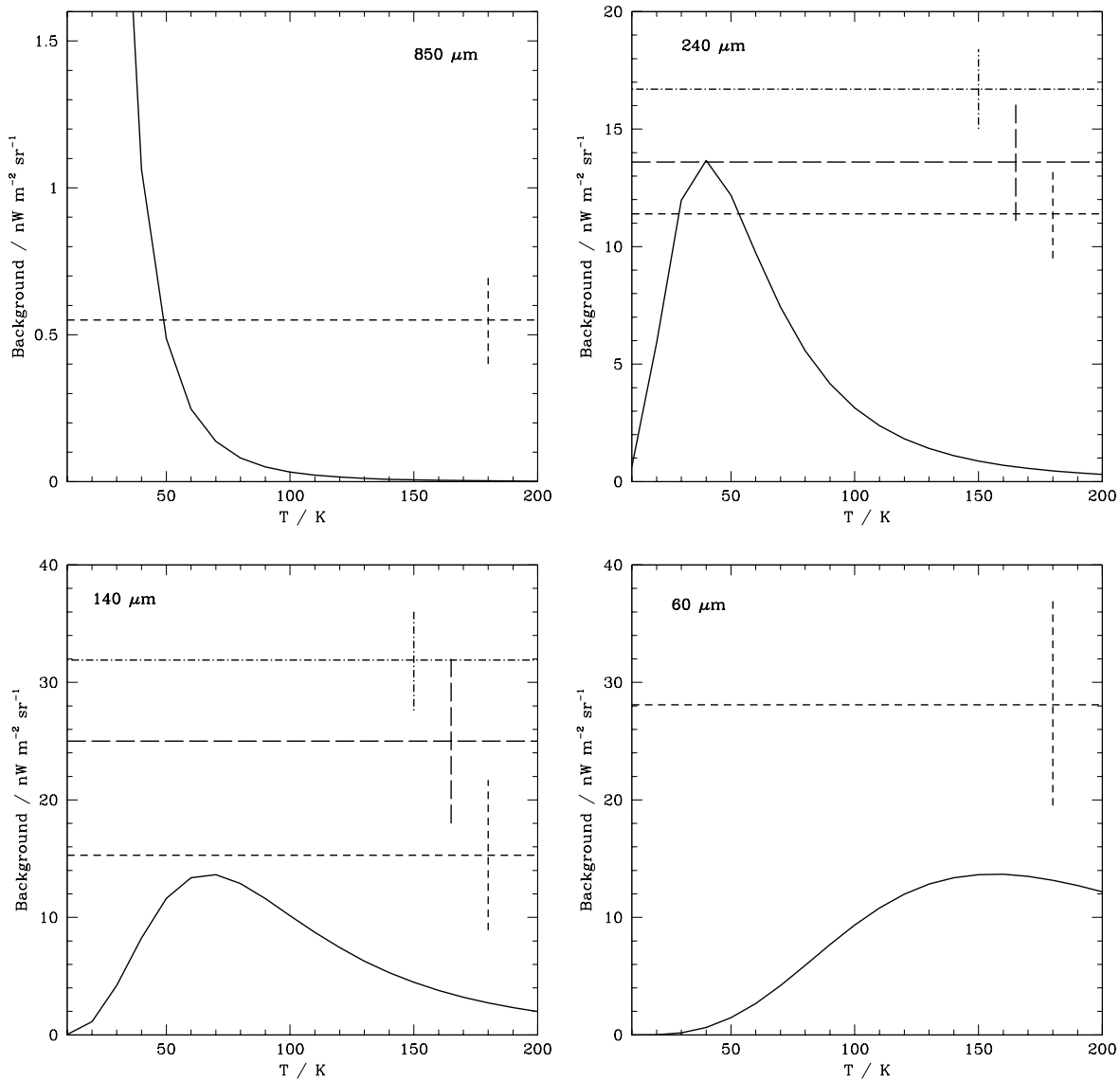, width=18.65cm}
\end{center}
{\vskip-5.7cm}
\caption{The contribution to the submillimetre and far-infrared 
extragalactic background light at various wavelengths 
from reprocessed starlight as a function of the mean flux-weighted 
temperature $T$ of the dust (solid lines). The amount of absorbed starlight 
is chosen to match the internal extinction corrections to the 
star-formation history derived by Steidel et al.\ (1999, Fig.\,9); 
these corrections also reproduce the observed value of the baryonic density
in stars $\Omega_* \approx 0.004$ (Pettini 1999), as we require if
the SCUBA sources are 
powered by AGNs rather than star formation. 
We assume that the shape of their luminosity function (Fig.\,5 of that 
paper) does not change with redshift, that the contribution to the 
star-formation rate from galaxies with
$M_{\rm AB}$(1700 \AA) $> -20$ or
$M_{\rm AB}$(1700 \AA) $< -23$
is negligible, and that the dust emissivity
index is 1.5. 
The dashed lines and errorbars 
represent observed backgrounds: top left -- Fixsen et al.~(1998); top right --
Schlegel et al.\ (1998; dot-dashed), Hauser et al.\ (1998;
long-dashed), Lagache et al.\ (1999; short-dashed); 
bottom left -- as top right; bottom right -- Finkbeiner et al.~(2000).} 
\end{minipage}
\end{figure*}

\section*{Acknowledgments} 

We thank Priya Natarajan, Max Pettini, and Dave Sanders
for helpful discussions,
and acknowledge PPARC and the Raymond \& Beverly
Sackler Foundation for financial support.

\vskip 8pt

\noindent{\bf APPENDIX A}
\vskip 3pt
\noindent
An alternative source of the 
submillimetre and far-infrared 
background radiation, and one which must 
contribute 
to some degree, is the reprocessed
radiation from young stars
that is absorbed and reradiated by dust in normal star-forming galaxies at
redshifts $0 \le z \le 4$ (Steidel et al.~1999).
The contribution to the background light at four different wavelengths
is shown in Fig.\,3 as a function of the temperature of the dust 
that is radiating 
the absorbed energy. If this temperature is between 50 and 100\,K, then 
much of the 140-$\mu$m background radiation intensity can be explained, 
while there is a negligible contribution to the 850-$\mu$m background, 
which can be produced almost entirely by the known population of 
more luminous
SCUBA sources. 
Dust temperatures of 50-100\,K are similar to those of the dust 
clouds that dominate the bolometric luminosity of 
Galactic star-forming regions (Wynn-Williams 1982). 
For $T=75$ K, the 850-$\mu$m flux of 
even 
a very luminous
$M_{\rm AB}$(1700 \AA) $= -23$ Lyman break
galaxy at $z=3$ would be only about 0.7 mJy, lower than the 
$> 2$\,mJy fluxes of the SCUBA sources. 
See also the analyses of Ouchi et al.\ (1999) and Peacock et al.~(2000).

\end{document}